\documentclass[12pt]{iopart}
\usepackage{graphicx,hyperref,iopams,feynmf}

\begin{document}

\title{Pairing in the iron arsenides: a functional RG treatment}

\author{Christian Platt, Carsten Honerkamp, and Werner Hanke}

\address{Institute for Theoretical Physics and Astrophysics, University of W\"urzburg, Am Hubland, 97074 W\"urzburg, Germany}
\ead{honerkamp@physik.uni-wuerzburg.de}
\begin{abstract}
We study the phase diagram of a microscopic model for the superconducting iron arsenides by means of a functional renormalization group. Our treatment establishes a connection between a strongly simplified two-patch model by Chubukov et al.\cite{Chubukov} and a five-band-analysis by Wang et al.\cite{d_h_lee}. For a wide parameter range, the dominant pairing instability occurs in the extended s-wave channel. The results clearly show the relevance of pair scattering between electron and hole pockets. We also give arguments that the phase transition between the antiferromagnetic phase for the undoped system and the superconducting phase may be first order.  
\end{abstract}

\maketitle

\section{Introduction}
The discovery of high-temperature superconductivity in the iron arsenides of the 1111\cite{hosono,56K} and the 122\cite{122ref} families with critical temperatures up to $T_c=56K$ has triggered enormous research efforts\cite{norman}. From the available experimental data and from density-functional theory\cite{LDA} it became clear rather quickly that the superconducting pairing is most likely due to electron-electron interactions. Furthermore, most evidence points to a weaker local correlation strength than in the high-$T_c$ cuprates\cite{anisimov}. Hence, weak coupling theories should actually work well in order to understand the phase diagram of these materials.

Very early, first random-phase-approximation (RPA) treatments indicated that the predominant pairing instability should be in the extended $s$-wave or $s\pm$-channel with a pairing amplitude of opposite sign on electron- and hole-like Fermi surfaces\cite{mazin}. However, a more complete theory should also account for vertex corrections and competing channels. Renormalization group (RG) methods go beyond RPA and sum all one-loop diagrams on equal footing up to infinite order. In two  inspiring papers\cite{d_h_lee}, the Berkeley group has already shown that the gross picture from the RPA is reproduced by the RG approach for two different descriptions\cite{raghu} of the band structure, and supports the extended-$s$ pairing state. However, their treatments of two- and five-band models also unveiled some differences between the predictions of the models, e.g. regarding the order parameter in the spin-density-wave (SDW) state. The complexity of the multi-band models indicated by these works, in particular due to the rich phase structure of the interactions in the Brillouin zone when the band language is used, definitely calls for other independent studies which show that the results do not depend critically on the details of the model and approximations used.   
Another, much simpler RG-like approach was recently presented by Chubukov and collaborators\cite{Chubukov}. They used a strongly simplified electronic structure with one hole-pocket and one electron pocket, both of circular shape. Furthermore, they assumed that the interactions near the Fermi surfaces only depend on the pocket and not on the precise location of a wavevector on the respective pocket. 
Also in this model, for quite general repulsive interactions, the leading pairing instability turns out to be in the extended $s$-wave channel. 

Our study below aims to interpolate between the Berkley group and Chubukov et al. Like in the simplified treatment by Chubukov et al.\cite{Chubukov}, we define the bare interactions in $\vec{k}$-space, only dependent on whether the wavevector is near the $\Gamma$ or near the $M$-point. In contrast to the work of the Berkeley group\cite{d_h_lee}, where the interactions are defined in real space, this will allow us to assess the effect of the wavevector-dependent matrix elements one gets from the transformation from real and orbital space into the band language. As the underlying band structure might be sensitive to details of the DFT scheme used to obtain it, this is an important check for the robustness of the results. It brings us closer to answering the question what the minimal model for the iron arsenide superconductivity is. Opposed to \cite{Chubukov}, in our functional RG additional wavevector-dependence around the Fermi surfaces can develop during the flows. This could possibly lead to different pairing states, but turns out to be rather uneventful in our calculations.
Furthermore, we use a more realistic band structure with two hole-like and two electron-like Fermi surfaces, such that we can study the effects of the nesting mismatch and doping in more detail than it was done in Ref. \cite{Chubukov}. This way we can get estimates for the energy scales of SDW and SC ordering, and get out some more qualitative trends.

This paper is organized as follows. In Sec. 2 we outline the functional RG approach as a tool for detecting instabilities in a Fermi liquid. In Sec. 3 we briefly summarize and comment on the two-patch model used by Chubukov. In Sec. 4 we present our extended modeling of the system, and in Sec. 5 we discuss the fRG results. Sec. 6 contains a summery and discussion.

\section{Functional RG}
The renormalization group is a well-known general concept to derive effective theories, e.g. at long length scales or for a low-energy window of a given many-particle system. For weakly coupled fermion systems, one is mainly interested in the effective interactions near the Fermi surface, as they contain information about possible symmetry-broken or non-Fermi-liquid ground states. Here, RG approaches to interacting fermions are more unbiased than  diagrammatic summations in a particular channel, as competing channels are treated on equal footing. The self-energy near the Fermi surface is also of high importance, but is typically harder to obtain. Moreover, many trends in the self-energy, like gap-openings are ``fore-shadowed'' in the effective interactions. To compute the effective  interactions near the Fermi surface, many recent works use the RG flow equations for the effective action or one-particle irreducible vertex functions\cite{wetterich,salmhofer,kopietz}, that avoid some complications of other straightforward adaptions of Wilsonian RG for interacting fermions\cite{shankar}. These RG schemes are commonly dubbed functional RG (fRG) as they aim at keeping as much of the wavevector- and possibly also frequency-dependence of the vertex functions, i.e. describe a flow of coupling functions rather than a flow of a finite number of coupling constants. 

We study a model described by an action, i.e.
\begin{equation}\begin{array}{ccl}
S (\psi,\bar\psi)& =&  \sum_{k,s} \,  \bar \psi_{k,s}
Q(k) \, \psi_{k,s} \\[2mm] &&
+ \frac{1}{2N}\, T^3 \, \sum_{k,k',q \atop
s,s'} V(k,k',k+ q)
\bar \psi_{k+q,s} \bar \psi_{k'-q,s'}
\psi_{k',s'}  \psi_{k,s} \, . \end{array}
\label{ham}  \end{equation}
Here $\bar \psi (k,s)$ and  $\psi (k,s)$ are Grassmann fields
representing fermions with wavevector $\vec{k}$, band index $b$, Matsubara
frequency $k_0$ (we write $k=(k_0, \vec{k},b)$) and spin projection
$s=\pm 1/2$. $Q(k)$ denotes the quadratic part of the action,
here given by
$
Q(k)= T \left[ -i k_0 + \epsilon (\vec{k})  \right] 
$. 
The RG is organized in the following way. One introduces a cutoff-function $C_\Lambda [\epsilon(\vec{k})]$
into the quadratic part of the action,
\begin{equation}  Q_\Lambda (k)
= T \sum_{i k_0 ,\vec{k}}  C_\Lambda [\epsilon(\vec{k})] \,
\bar{\psi}_k  \, \left[ -i k_0 + \epsilon (\vec{k})\right]\, \psi_k  \, .
\end{equation}
$C_\Lambda [\epsilon(\vec{k})]$ is very large for $|\epsilon (\vec{k})| \le
\Lambda$ and $C_\Lambda [\epsilon(\vec{k})]=1$ for $|\epsilon (\vec{k})| > \Lambda$  such that modes below $\Lambda$ are not integrated over 
in the functional integral.
In practice, one mainly needs the inverse $C^{-1}_\Lambda
[\epsilon(\vec{k})]$, which is chosen as a smoothened-out step
function for numerical treatments. The Green's function as the inverse of $Q_\Lambda$ is suppressed  for modes with  $|\epsilon (\vec{k})| \le \Lambda$. Here we neglect any self-energy corrections. 
For spin-rotational and U(1) invariance, the
antisymmetric four-point interaction vertex for incoming particles $k_1,s_1$,
$k_2,s_2$ and outgoing particles $k_3,s_3$, $k_4,s_4$ (the quantum
numbers of particle 4 are dictated by the conservation laws, except for the band index $b_4$ of the second outgoing particle) can
be expressed in terms of a spin-independent coupling function $V(k_1,k_2,k_3,b_4)$ which still carries the symmetries of the lattice\cite{salmhofer}.
The central RG differential equation governing the flow of the effective interactions
$V_\Lambda (k_1,k_2,k_3,b_4)$ is\cite{salmhofer}
\begin{eqnarray}
\frac{d}{d\Lambda} V_\Lambda (k_1,k_2,k_3,b_4) =    {\cal T}_{PP,\Lambda} + {\cal T}^d_{PH,\Lambda} +
{\cal T}^{cr}_{PH,\Lambda}, \label{vdot} \end{eqnarray} 
with the one-loop particle-particle contributions ${\cal T}_{PP,\Lambda}$ and the two different particle-hole channels ${\cal T}^d_{PH,\Lambda}$ and ${\cal T}^{cr}_{PH,\Lambda}$, where
\begin{eqnarray}
\lefteqn{  {\cal T}_{PP,\Lambda} (k_1,k_2;k_3,k_4) = }\nonumber\\
 - &&\int dk \sum_{b'}\, V_\Lambda ( k_1,k_2,k, b' ) \, L_{b,b'}(k,-k+k_1+k_2) \, V_\Lambda
(k,-k+k_1+k_2 ,k_3,b_4) \label{PPdia}
\\[0mm]
 \lefteqn{ {\cal T}^d_{PH,\Lambda} (k_1,k_2;k_3,k_4) =}  \nonumber \\
-&& \int dk\,\sum_{b'} \biggl[ -2 V_\Lambda ( k_1,k,k_3,b' ) \, L_{b,b'}(k,k+k_1-k_3) \,
V_\Lambda (k+k_1-k_3,k_2,k,b_4) \nonumber
 \\ && \qquad + 
V_\Lambda (k_1,k,k+k_1-k_3,b') \, L_{b,b'}(k,k+k_1-k_3) \, V_\Lambda (k+k_1-k_3,k_2,k,b_4)
 \nonumber  \\ && \qquad + 
V_\Lambda ( k_1,k,k_3,b') \, L_{b,b'}(k,k+k_1-k_3) \, V_\Lambda (k_2,k+k_1-k_3,k, b_4)
\biggr]\label{PHddia}
\\[0mm]
\lefteqn{  {\cal T}^{cr}_{PH,\Lambda}(k_1,k_2;k_3,k_4) =} \nonumber \\
-&& \int dk \sum_{b'} \, V_\Lambda (k_1,k+k_2-k_3,k,b')  \, L_{b,b'}(k,k+k_2-k_3) \,
V_\Lambda (k,k_2,k_3,b_4 )\label{PHcrdia}
\end{eqnarray}
In these equations, $\int dk = T \sum_{ik_0} \int \frac{d^2k}{(2\pi)^2}$, and the band index $b'$ of the second internal line is to be summed over, while its wavevector and frequency are determined by conservation laws, e.g. as $-k+k_1+k_2$ in Eq. \ref{PPdia}. 
The product of the two internal lines in the
one-loop diagrams is
\begin{equation} \label{e13}
L_{b,b'}(k,k') = S_\Lambda (k) G^{(2)}_{\Lambda} (k') + G^{(2)}_{\Lambda} (k)
S_\Lambda (k'),  \, \end{equation} 
with the band indices $b$, $b'$ belonging to $k$ and $k'$ and the so-called single-scale propagator
\begin{equation}
S_\Lambda (k) = - G^{(2)}_{\Lambda} (k) \left[ \frac{d}{d\Lambda} Q_\Lambda
(k) \right] \, G^{(2)}_{\Lambda} (p) \, .
\end{equation}
Here, $G^{(2)}_{\Lambda} (k)$ denotes the full Green's function at the RG scale $\Lambda$, but as stated above in this work we do not include self-energy corrections. Equations (\ref{vdot}) can be integrated numerically for initial values corresponding to the bare interactions. At low $\Lambda$ and low temperatures $T$, the flow will typically lead to strong coupling, i.e. a certain part of the $V_\Lambda (k_1,k_2,k_3)$ will flow to values much larger than the bandwidth. Then the flow has to be stopped as the approximations (neglect of selfenergy, truncation of the hierarchy of the flow equations) will break down. Nevertheless, the scale where the flow to strong coupling occurs gives an upper bound for the energy or temperature scales of ground state ordering. Most importantly, the analysis of which channel diverges most severely indicates the type of ground state state ordering. 

The fRG equations stated above are valid at any temperature. Estimates for critical temperatures of long-range-ordered phases can be obtained  by determining the temperature above which the flow remains finite down to zero scale. However, in standard cases, this value for $T_c$ can (up to a factor of order one) also be inferred from the energy scale for the flow to strong coupling $\Lambda_c$ at $T=0$. For instance, for a reduced Cooper problem with attractive interactions, we would get $T_c= 1.14 \Lambda_c$ where $\Lambda_c = W \exp (-1/g)$ with the bandwidth $W$ and the dimensionless coupling constant $g$. Hence, in the following we only compute $\Lambda_c (T=0)$. Furthermore, the approximate flow equations used here do not capture the destruction of long-range order in two dimensions by soft modes in the case of continuous symmetries. Yet, in the present study, this is not a problem, as in the ``back of our minds'' we actually want to study quasi-2D systems with some coupling in the third direction. There this destructive mechanism by Goldstone-modes is not effective, and spontaneous breaking of continuous symmetries is indeed possible.  

\section{Two-patch model}
In this section, we first describe the most drastic but still meaningful simplification of the functional RG for the pnictide problem. Photo-emission experiments and density functional theory on the iron pnictides reveal (at least) 4 Fermi surfaces in the Brillouin zone\cite{norman}. Using the unit cell with two iron and two arsenic atoms, one has two nearly circular hole-like Fermi surfaces around the $\Gamma=(0,0)$-point and two more deformed electron-like Fermi surfaces around the $M=(\pi,\pi)$-point (for details see Fig. \ref{dispersion}). The simplified model by Chubukov et al.\cite{Chubukov} reduces these Fermi surfaces to two circles, one hole-like Fermi circle around the $\Gamma$- and one electron-like circle around the $M$-point. In a further reduction step, the interactions between the quasiparticles states near these Fermi surfaces were taken to depend only on whether the excitation occurs near the hole-like or the electron-like Fermi surface. For the particle-hole-symmetric case this leads to 4 independent couplings $g_1, \dots ,g_4$ which are indicated in Fig. \ref{patching}(b). In comparison with Ref. \cite{Chubukov} we have interchanged $g_1$ and $g_2$.  The RG flow, upon integrating out the single-particles state with decreasing energy scale $\Lambda$, gives rise to the equations\cite{Chubukov}:
\begin{eqnarray} \dot g_1 = 2g_1(g_2 - g_1) \\
\dot g_2 = g^2_2 + g^2_3 \\
\dot g_3 = -2g_3g_4+2g_3(2g_2-g_1)\\
\dot g_4 = -g^2_3-g^2_4.
\end{eqnarray}
Here the notation $\dot g_i = \frac{\Lambda}{\rho(0)} \frac{dg_i}{d\Lambda}$ is used, and $\rho(0)$ is the density of states at the Fermi level per Fermi circle and spin component. 
The general flow for repulsive bare interactions leads to strong coupling, i.e. the $g$'s diverge at a nonzero scale $\Lambda_c$. $g_2$ and $g_3$ flow to $+\infty$ while $g_4$ goes to $-\infty$. $g_1$ diverges less strongly. If one now considers how the effective coupling constants for various symmetry breaking channels flow, one finds that several channels develop instabilities at the same critical scale. For example the SDW coupling constant $g_2+g_3$ diverges, as does the coupling constant in the $s\pm$-pairing channel, i.e. $g_3-g_4$, and some other combinations. Since for repulsive interactions, the coupling constant in the SDW channel is always larger than in the pairing channel, Chubukov et al. interpret this flow as indicative of a SDW instability. 
While we believe that this is physically correct, it should be noted that in one-dimensional systems like two-leg Hubbard ladders one finds similar multi-channel instabilities with a joint divergence of SDW and unconventional pairing channels. In these systems ,however, the ground state does not develop SDW order but maintains a spin gap, and no power-law correlations occur.
In physical terms, the competition of several ordering tendencies does not result in a clear ``winner'' with some kind of (quasi-)long-range order, but in a gapped spin-liquid-like ground state that embodies various types of strong short-range correlations.  

In the efforts to understand the one-band Hubbard model on the two-dimensional square lattice near half-filling, many authors have studied  the so-called two-patch model\cite{2patchrefs}. Here one assumes that the Fermi surface is near the van Hove points at $(0,\pi)$ and $(\pi,0)$, and that only these $\vec{k}$-space regions are important for the breakdown of the Fermi liquid one is interested in. Again, in the standard treatment, the interactions are taken to depend only on the patch index, and not on the precise location of the wavevector inside the patch. This way one has to consider again four coupling constants $g_1$ to $g_4$, defined very analogously to the pncitide toy model described above. It turns out that the RG equations for the one-band Hubbard model in the two-patch approximation are just the ones of the pnictide two-patch model described above, with the slight difference that now the density of states actually diverges at the Fermi level placed at the van Hove energy. However, on this level of approximation, the density of states only regulates the magnitude of the critical scale, and does not change the interplay of the various couplings and symmetry breaking channels. Again the flow is  given by $g_2$ and $g_3 \to \infty$, while $g_4 \to - \infty$, and AF-SDW, $d$-wave pairing and $d$-density wave susceptibilities diverge together.  In the cuprates and in analogy with the ladder problem mentioned above, this picture was interpreted as an indication for a complex ``mother state'' with several competing channels embedded, with possible relevance to the pseudogap of the underdoped cuprates\cite{hsfr,laeuchli}. Note that very recently, the interplay of several ordering tendencies in the model and approximation by Chubukov et al.\cite{Chubukov} has been understood\cite{podolsky,chubukov09} as a SO(6)-symmetric fixed point of the RG flow with degeneracy of pairing, SDW and orbital density wave channels. It would be very interesting to learn more about properties like quasiparticle spectra in the vicinity of this unstable fixed point.

For the pnictides, there is at present no indication to argue for a complex spin-liquid state. Hence one might ask if the minimal two-patch model overestimates the channel coupling and suggests a more sophisticated picture than what is actually found in the experimental system. In terms of the SO(6)-symmetry\cite{podolsky}, the flow seems to veer away from the fixed point relatively early and is instead attracted by other fixed points with only one type of ordering.

For the pnictides, there is at present no indication to argue for a complex spin-liquid state. Hence one might ask if the minimal two-patch model overestimates the channel coupling and suggests a more sophisticated picture than what is actually found in the experimental system. Indeed, in the following, we will show that a generalization to more than just four coupling constants results, depending on the band filling, in two quite distinct regimes with only one predominant instability. Furthermore, one could ask if the simplification to the 4 coupling constants with no dependence is legitimate or if additional, possibly relevant wavevector-dependences arise in the flow. Below we will see that is not the case. 
 
\section{Model and FRG}
To model the band structure of the iron pnictides, we employ a four-band model in the folded Brillouin zone (BZ) proposed by Korshunov and Eremin \cite{korshunov}. The free Hamiltonian reads as
\begin{equation}\label{modelhamilton}
H_0=-\sum_{\vec{k},i,\sigma}\epsilon^i n_{i,\sigma}(\vec{k}) - \sum_{\vec{k},i,\sigma}
t_{\vec{k}}^id^{\dagger}_{i\sigma}(\vec{k})d^{\phantom{\dagger}}_{i\sigma}(\vec{k}),
\end{equation}
with band indices $i=\alpha_1,\alpha_2,\beta_1,\beta_2$, onsite energies $\epsilon^i$ and an electronic dispersion $t_{\vec{k}}^i$ given by
\begin{eqnarray}\nonumber
t_{\vec{k}}^{\alpha_1,\alpha_2}&&=t_1^{\alpha_1,\alpha_2}(\cos{k_x}+\cos{k_y}) + t_2^{\alpha_1,\alpha_2}
\cos{k_x}\cos{k_y},\\\nonumber
t_{\vec{k}}^{\beta_1,\beta_2}&&=t_1^{\beta_1,\beta_2}(\cos{k_x}+\cos{k_y}) + t_2^{\beta_1,\beta_2}
\cos{\frac{k_x}{2}}\cos{\frac{k_y}{2}}.
\end{eqnarray}
Using the condensed notation $(\epsilon^i,t_1^i,t_2^i)$, we assume parameter values of
$(-0.60,0.30,0.24)$ and $(-0.40,0.20,0.24)$ for the $\alpha_1$ and $\alpha_2$ bands, respectively,
and $(1.70,1.14,0.74)$, $(1.70,1,14,-0.64)$ for the $\beta_1$ and $\beta_2$ bands (all values are in units of $eV$). 
\begin{figure} 
\centering
{\includegraphics[width=0.9\textwidth]{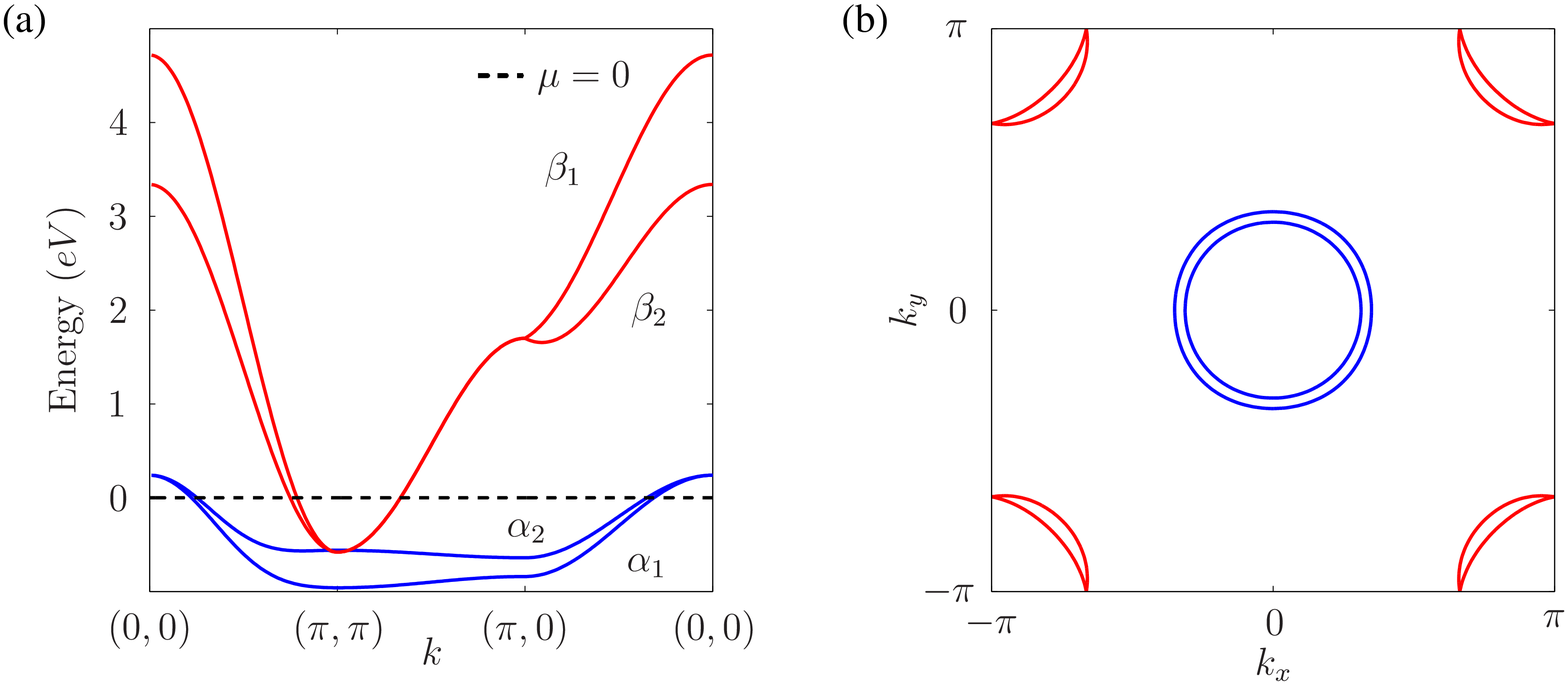}}
\caption{(color online) (a) Energy dispersion along the main symmetry points of the first BZ. (b) Fermi surface topology at half-filling consisting of two hole pockets (blue) around $\Gamma=(0,0)$ and two electron pockets (red) near $M=(\pi,\pi)$ in the folded BZ.}
\label{dispersion}
\end{figure}
The doping is controlled by adding a chemical potential term to Eq. (\ref{modelhamilton}). A properly adjusted $\mu$ then yields
two circular hole pockets around $(0,0)$ and two elliptic electron pockets near $(\pi,\pi$) in the folded Brillouin zone 
as depicted in Fig. \ref{dispersion}(b). 
For the undoped situation, the areas of electron- and hole-pockets are the same, and the total electron number per spin projection and unit cell is $n_s=2$, or $n=4$, if we sum over the spins.
As we have 2 Fe ions in the unit cell, a doping of $x$ electrons per Fe for the 1111 compounds gives a band filling $n=4+2x$.
The band dispersion of the model Hamiltonian (\ref{modelhamilton}), as well as its Fermi surface topology at half-filling ($\mu=0$), is shown in Fig. \ref{dispersion}. Here, it is clearly visible that hole and electron pockets are almost perfectly nested for $Q=(\pi,\pi)$ which causes a logarithmic divergence in the particle-hole channel and, thus, drives the $(\pi,\pi)$-SDW instability.

\begin{figure} 
\centering
{\includegraphics[width=0.9\textwidth]{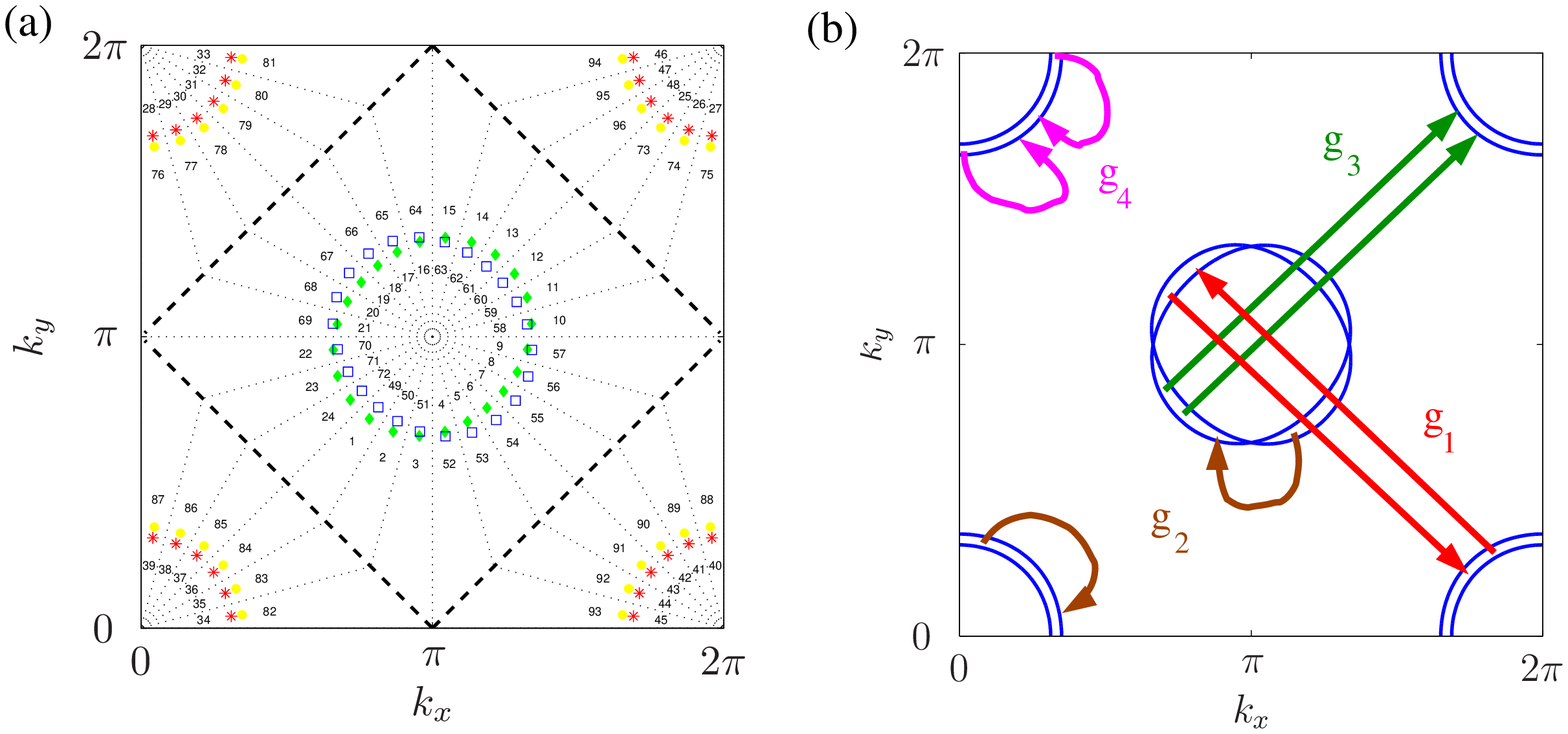}}
\caption{\label{patching}(color online) (a) Discretization of the Brillouin zone with a 96-patch system. (b) The four relevant scattering processes near the FS with coupling constants $g_1$ (interband), $g_2$ and $g_4$ (intraband) as well as an interband pair hopping $g_3$. For each process $g_i$, the two arrows denote the scattering of quasiparticles with fixed spin projection $s$ on one line and $s'$ on the other line.}
\end{figure}

Next, we include electron-electron interactions consisting of intra- ($g_4$) and interband interactions ($g_1$ and $g_2$) as well as an interband pair hopping ($g_3$) in direct analogy to the two-patch model by Chubukov et al.\cite{Chubukov} discussed in the previous section.
Summing up these contributions, the interaction part can be written as 
\begin{eqnarray}\nonumber
H_I&&=
\frac{g_1}{2}\sum_{\vec{k},\vec{k'},\vec{q}\atop \sigma , \sigma'}d^{\dagger}_{\alpha\sigma}(\vec{k} + \vec{q})d^{\dagger}_{\beta\sigma'}(\vec{k}'-\vec{q})d^{\phantom{\dagger}}_{\alpha\sigma'}(\vec{ k}')d^{\phantom{\dagger}}_{\beta\sigma}(\vec{k})+ h.c.\\\nonumber
&&+\frac{g_2}{2}\sum_{\vec{k},\vec{k'},\vec{q}\atop \sigma , \sigma'}d^{\dagger}_{\alpha\sigma}(\vec{k} + \vec{q})d^{\dagger}_{\beta\sigma'}(\vec{k}'-\vec{q})d^{\phantom{\dagger}}_{\beta\sigma'}(\vec{ k}')d^{\phantom{\dagger}}_{\alpha\sigma}(\vec{k}) + h.c. \\\label{interaction}
&&+\frac{g_3}{2}\sum_{\vec{k},\vec{k'},\vec{q}\atop \sigma , \sigma'}d^{\dagger}_{\alpha\sigma}(\vec{k} + \vec{q})d^{\dagger}_{\alpha\sigma'}(\vec{k}'-\vec{q})d^{\phantom{\dagger}}_{\beta\sigma'}(\vec{ k}')d^{\phantom{\dagger}}_{\beta\sigma}(\vec{k}) + h.c.\\\nonumber
&&+\frac{g_4}{2}\sum_{\vec{k},\vec{k'},\vec{q}\atop \sigma , \sigma'}d^{\dagger}_{\alpha\sigma}(\vec{k} + \vec{q})d^{\dagger}_{\alpha\sigma'}(\vec{k}'-\vec{q})d^{\phantom{\dagger}}_{\alpha\sigma'}(\vec{ k}')d^{\phantom{\dagger}}_{\alpha\sigma}(\vec{k})\\\nonumber
&&+\frac{g_4}{2}\sum_{\vec{k},\vec{k'},\vec{q}\atop \sigma , \sigma'}d^{\dagger}_{\beta\sigma}(\vec{k} + \vec{q})d^{\dagger}_{\beta\sigma'}(\vec{k}'-\vec{q})d^{\phantom{\dagger}}_{\beta\sigma'}(\vec{ k}')d^{\phantom{\dagger}}_{\beta\sigma}(\vec{k}),
\end{eqnarray}
where the sum over all band index combinations with $\alpha\in\{\alpha_1,\alpha_1\}$ and $\beta\in\{\beta_1,\beta_2\}$ is implicitly assumed. 
For the numerical handling of the RG flow in Eq. (\ref{vdot}), we discretize the Brillouin zone into $N$ segments (patches) as shown in Fig. \ref{patching}(b), in generalization to the fRG-treatments of the one-band Hubbard model\cite{zanchi,hsfr}. Each segment is confined within two neighboring dotted lines running from $(\pi,\pi)$ to one of the corner points $(\pi\pm\pi,\pi\pm\pi)$. If we consider an arbitrary momentum $\vec{k}_i$ and a freely chosen band index $b_i$, the point ${\bf k}_{F}(\vec{k}_i,b_i)$ is defined as the intersection of the angle bisector lying in the same segment as $\vec{k}_i$ with the FS sheet corresponding to band index $b_i$. This provides $4N$ different discretization points pictured as colored markers in Fig. \ref{patching}(b). Points 1-24 run counter-clockwise around the first electron pocket, points 25-48 around the first hole pocket, points 49-72 around the second electron pocket and points 73-96 from the second hole pocket.
Since the leading part of the flow is given by the coupling function on the Fermi surface and at zero frequencies \cite{salmhofer}, we approximate the effective coupling $V_{\Lambda}(k_1,k_2,k_3,b_4)$ by
$V_{\Lambda}(k_1,k_2,k_3,b_4)=V_{\Lambda}(\tilde{k}_1,\tilde{k}_2,\tilde{k}_3,b_4)$, where we used $\tilde{k}_i=(0,{\bf k}_{F}(\vec{k}_i,b_i),b_i)$. This implies that each momentum $\vec{k}_{1,\ldots 3}$ can be varied within its segment, while leaving the coupling function $V_{\Lambda}(k_1,k_2,k_3,b_4)$ equal to $V_{\Lambda}(\tilde{k}_1,\tilde{k}_2,\tilde{k}_3,b_4)$. 
Due to this approximation, it is useful to decompose the $k$-space integration in Eq. (\ref{vdot}) into a radial part with
a constant coupling in the integrand and a subsequent averaging over all segments.
Starting from an energy scale $\Lambda_0$ of the order of the bandwidth and $V_{\Lambda_0}(k_1,k_2,k_3,b_4)$ determined by the bare interactions $g_1,\ldots,g_4$, the flow equation can now be integrated down to a critical scale $\Lambda_c$ where certain parts of   
$V_{\Lambda_c}(k_1,k_2,k_3,b_4)$ start to diverge. Similar to the RG flow of the effective coupling function, we also compute the flow of certain susceptibilities in order to have a clear criterion for the leading instability. Therefore, we consider the coupling of external fields $\Phi_{sc}$ and $\Phi_{sdw}$ to fermionic bilinears written as
\begin{eqnarray}\nonumber
H_{sc} &= \Phi_{sc}\sum_{\vec{k},s\atop a,a'}h_{sc}(\vec{k})
\left(d^\dagger_{{\vec{k}},a,s}d^\dagger_{-\vec{ k},a',-s} - d^\dagger_{\vec{k},a,-s} d^\dagger_{-\vec{k},a',s}\right)\\\nonumber
H_{sdw}&= \Phi_{sdw}\sum_{\vec{k},s\atop a,a'}h_{sdw}(\vec{k})\left(d^\dagger_{\vec{k} + \vec{Q},a,s}d_{\vec{k},a',s} - d^\dagger_{\vec{k} + \vec{Q},a,-s}d_{\vec{k},a',-s}\right),
\end{eqnarray}
with band indices $a,a'$ running over the four Fermi surfaces. The form factor is taken to be band-independent, but in the numerical implementation for the pairing case, $a$ and $b$ are restricted to be either both on electron-pockets or both on the hole-pockets. For the SDW channel, one particle at a hole-like Fermi surface and one in at an electron-like Fermi surface. 

These couplings are renormalized by one-loop corrections that involve the scale-dependent interactions.  In the following, we will only consider static external couplings. Then we can set the frequency flowing through these loop diagrams equal to zero. This amounts to $Q_0=0$ for the SDW case and zero total incoming frequency for the pairing channel. Using the notation $k=(0,\vec{k},a)$ with $a'$ for the second external band index, and $k'=(i k'_0 ,\vec{k}',b)$ and $b'$ for the second internal band index, the one-loop renormalization of these couplings is obtained by 
\begin{eqnarray}\nonumber
\frac{d}{d\Lambda} h_{sdw}(\vec{k}) &&= -\int dk \, h_{sdw}(\vec{k}')V_{\Lambda}({ k},{ k'},{ k'+Q},a') L_{b,b'} ({ k'},{ k' + Q})\\\nonumber
\frac{d}{d\Lambda} h_{sc}(\vec{k}) &&= -\int dk'\, h_{sc}(\vec{k}')V_{\Lambda}({ k},-{ k},{ k'},b') L_{b,b'}({ k'},-{ k'}),
\end{eqnarray}
with $L_{b,b'}(\cdot,\cdot)$ defined in Eq. (\ref{e13}) and the summation over all involved internal band indices implicitly assumed.
The one-loop flow of the susceptibilities is calculated in a similar manner, i.e.
\begin{eqnarray}\nonumber
\frac{d}{d\Lambda} \chi_{sdw} &&= -\int d\vec{k}\ h_{sdw}(\vec{k})L_{bb'} ({ k},{ k + Q})h_{sdw}(\vec{ k}+\vec{Q}),\\\nonumber
\frac{d}{d\Lambda} \chi_{sc} &&= -\int d\vec{k}\ h_{sc}(\vec{k}) L_{bb'} ({ k},-{ k})h_{sc}(-\vec{k}).
\end{eqnarray}
 As initial conditions for the couplings, we choose $h_{sdw}(\vec{k})=1$ for the spin coupling and $h^s_{sc}(\vec{k})=1$, $h^{s\pm}_{sc}(\vec{k})=\cos(k_x) + \cos(k_y)$ or $h^d_{sc}(\vec{k})=\cos(k_x) - \cos(k_y)$ for the different pairing channels. The initial values for the susceptibilities are zero. It becomes obvious that certain classes of coupling functions, as for instance $V(k,-k,k')$ and $V(k,k',k'+Q)$, drive the pairing or, respectively, the spin susceptibility and hence indicate a possible instability in this channel. 

\section{Flows to strong coupling and phase diagram}
In the following, we calculate the RG-flow of the coupling function $V(k_1,k_2,k_3,b_4)$ and of the susceptibilities described in Sec. 4. Using a 96 point discretization scheme, we examine the results for several fillings in the electron and hole doping regime.
As bare interaction parameters, let us first choose $g_1=g_2=g_3=g_4=0.4eV$. The flow is stopped if one element of  $V(k_1,k_2,k_3,b_4)$
exceeds $25eV$. 
\begin{figure} 
\centering
{\includegraphics[width=0.78\textwidth]{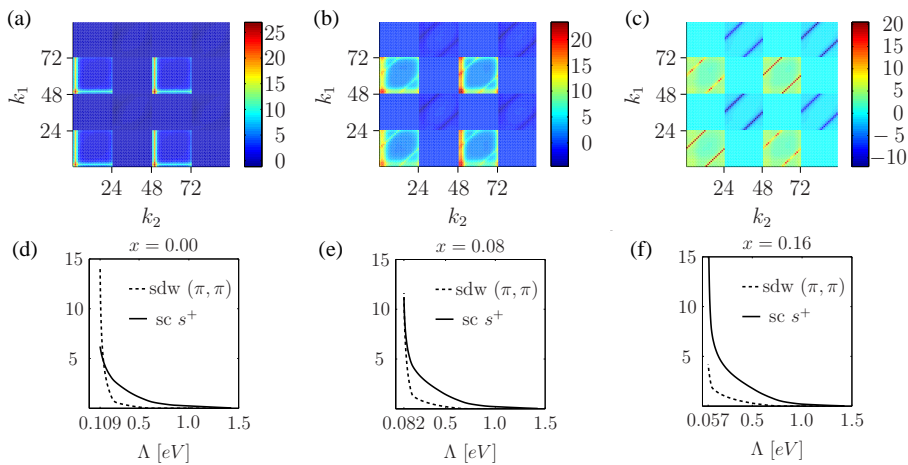}}
\caption{\label{flowres}(color online) (a-c) The coupling functions $V_{\Lambda}(k_1,k_2,k_3,b_4)$, discretized at 96 points according to Fig. \ref{patching}(a) with the first outgoing wave vector $k_3$ fixed at point 26 and the band index of the fourth momentum set to 2 corresponding to the first hole pocket with points 25-48. (a), (b) and (c) are calculated for electron dopings of $x=0.0$, $x=0.08$ and $x=0.16$, respectively. The colorbars indicate the values of the couplings and for $k_2= 2$, and $\vec{k}_3-\vec{k}_2=(\pi,\pi)$. (d-f) Flow of the $s^\pm$-wave (solid) and $(\pi,\pi)$-spin susceptibility (dashed).}
\end{figure} 
Starting with the half-filled case, the coupling function $V(k_1,k_2,k_3,b_4)$ in Fig. \ref{flowres}(a) with the third momentum fixed to point 26 and the fourth band index set to 2, clearly shows a vertical diverging feature. It corresponds 
to couplings $V(k,k',k'+Q)$ with a fixed momentum relation of $Q=(\pi,\pi)$ between the second and third momentum. 
As discussed in Sec. 4, such couplings with fixed wavevector-transfer drive the spin susceptibility with ordering momentum $Q$ and, thus, indicate a SDW instability. Correspondingly, the SDW instability in Fig. \ref{flowres}(d) diverges steeply at the critical scale, leaving behind the extended $s$-wave pairing susceptibility. 

Note that the instability described here only signals that the magnetic moments in neighbored unit cells are oriented in opposite directions.  In order to infer information of the ordering within the unit, we would have to define a real space picture, i.e. to determine the orbital and sublattice site content of the bands of our dispersion. The dominance of the SDW-susceptibility is still found at small nonzero doping, but the critical scale goes down gradually with doping and the pairing channel becomes a competitor, as seen in Fig. \ref{flowres}(b) and (e).

For electron doping of $x=0.16$ in Fig. \ref{flowres}(c), the leading divergence has changed significantly. Here, we find diagonal features which signify an intense scattering $V(k,-k,k',b_4)$ of total-wavevector-zero Cooper pairs. Moreover, there is a clear sign reversal in the pair scattering at point $k=24$, $k=48$ and $k=72$, which represents the change of $k$ from a hole-like to an electron-like FS and $vice$ $versa$. 
\begin{figure} 
\centering
{\includegraphics[width=.9\textwidth]{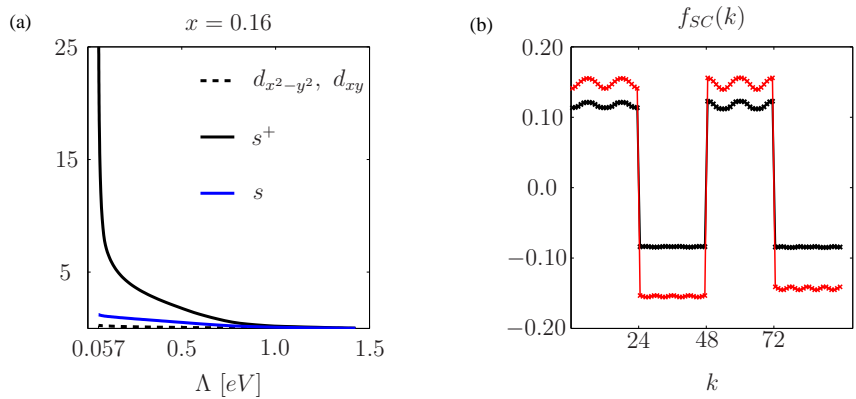}}
\caption{\label{scwave}(color online) (a) Flow of the $d-$wave (dashed), $s$-wave (blue solid) and $s^\pm$-wave (black solid) pairing susceptibility at $16\%$ electron doping. (b) The form factor $f_{SC}(k)$ (black) determined by the eigenvector corresponding to the largest eigenvalue in magnitude of the matrix $V_{\Lambda_c}(k,-k,q)$. For comparison, we also plot $0.1(\cos(k_x) + \cos(k_y))$ (red) evaluated at the discretization points in Fig. \ref{patching}.}. 
\end{figure} 
If we consider $(\cos(k_x) + \cos(k_y))$ as the simplest lattice harmonic generating this sign change, the pair scattering is properly fitted by $V(k,-k,k',-k')=-(\cos(k_x) + \cos(k_y))(\cos(k'_x) + \cos(k'_y))$ with the incoming pairing partners $k,-k$ on the same band $a$ and the outgoing pair on band $b$. 
As can be seen in Fig. \ref{flowres} c) there is also a growth of the scattering of interband pairs, e.g. with the partners on the two different electron pockets, that is however subleading, as the total wavevector of such pairs adds up to a small nonzero value. This difference is due to our patching scheme. 
As a word of caution, we cannot exclude that the interaction of zero-total momentum interband pairs might become as strong as the one of intraband pairs. This issue will be scrutinized in further studies.

An alternative analysis of the pairing symmetry can be obtained by optimizing the coupling strength functional \cite{raghu} which in our case reduces to an eigenvalue problem for the pairing matrix $V_{pair}(k,k')=V(k,-k,k',b_4)$. Here, the eigenvalues and eigenvectors of $V_{pair}$ represent
the pairing strength and the corresponding form factor. For the largest eigenvalue in magnitude, we depicted the corresponding 
eigenvector in Fig. \ref{scwave}(b) and also calculated the flow of the pairing susceptibility for different form factors and intraband pairs in
Fig. \ref{scwave}(a). 
 
These results clearly prove an extended $s$-wave pairing symmetry in accordance with the previous RG studies
\cite{d_h_lee,Chubukov}. Note also, e.g. by looking at the plots of the low-scale interactions in Fig. \ref{flowres}, that we find very clean SDW at $x=0$ or pairing instabilities for $x=0.16$, respectively. In these plots, only a single feature, i.e. either the vertical lines plus the related weaker horizontal feature of an effective long-range SDW interaction, or the diagonal feature corresponding to the pairing instabilities, are visible. This shows that the multi-channel instability found in the two-patch analysis overestimates the channel coupling, and Chubukov et al.\cite{Chubukov} were correct in their interpretation as a SDW instability. Furthermore, near the critical doping where the pairing susceptibility begins to grow more strongly than the SDW susceptibility, there is little overlap between SDW and pairing features in the effective interactions. This is unlike the situation in the one-band Hubbard model in the saddle point regime\cite{hsfr} and is well compatible with a simple change in the ground state from one type of order to the other without any exotic intermediate phases.  

From the pairing form factor in Fig. \ref{scwave}(b) we can infer that the pairing is equally strong on all Fermi surface sheets. This is a difference to conclusions drawn from photoemission experiments that showed a smaller gap on one of the hole pockets\cite{ding}. Our results indicate that this disparity may be due to a more detailed structure of the interactions near the Fermi surfaces, which is not included here. As another difference to microscopic studies based on more realistic five-band models\cite{d_h_lee,maier} is the mild variation of the pairing amplitude on the electron pockets. In the work by Graser et al.\cite{maier}, the pairing even changes sign on the electron pockets, however in a way preserving the extended $s$-symmetry. Again we suspect that the matrix elements from the transformation from the full band structure to the band picture included in these works and not in ours is responsible for this difference.

Pairing components different from the extended $s$-wave channel flow much more weakly. In particular, the $d$-wave component found in several studies \cite{d_h_lee,maier} is practically invisible in the effective couplings, and its susceptibility only shows a mild rise at low scales. 
The RG-flow for the hole doped case is not depicted here since the findings are quite similar as in the electron doped regime.

\begin{figure} 
\begin{center} 
\includegraphics[width=.9\textwidth]{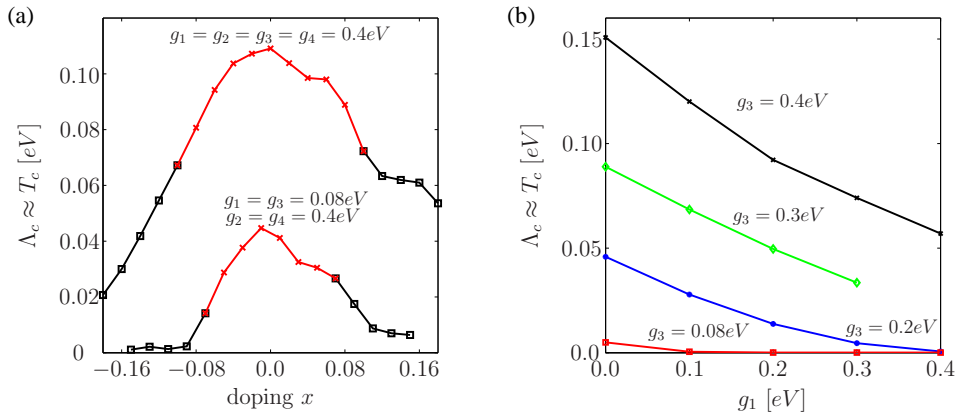}
\end{center}
\caption{\label{phasediag}(color online) (a) Phase diagram for different initial interactions. The crosses (red) and squares (black) indicate the SDW and superconducting transition, respectively. For the upper line we used $g_1=g_2=g_3=g_4=0.4eV$ as initial interaction parameters and for the lower one $g_1=g_3=0.08eV$, $g_2=g_4=0.4eV$. (b) Effective coupling $V_{\Lambda}(k_1,k_2,48,4)$
near the critical energy scale $\Lambda_c$ for initial interactions $g_1=g_3=0eV$, $g_2=g_4=0.4eV$ and electron doping $x=0.16$.}
\end{figure}

Next, we determine the critical scale $\Lambda_c$ as a function of doping and examine which of the susceptibilities $\chi_{SDW}$ and $\chi^{s\pm}_{sc}$ diverges first. The results are summarized in the phase diagram of Fig. \ref{phasediag}. Near $x=0$, the critical scale is highest, while it decays away from $x=0$. On the tails, we have marked the regions where the extended $s$-wave pairing susceptibility diverges more strongly than the SDW component. Even though the absolute value of critical dopings and critical scales strongly depend on the values of the initial interactions, we find an interesting trend in the data: If we compare the critical scales for the pairing instability at doping $\pm x$, i.e. the same absolute value of hole or electron doping, the critical scale is higher on the electron-doped side. In the experiments, although more factors might influences the situation as well, the higher $T_c$'s are likewise measured on the electron-doped side. In our theory, this trend is not easily understood in terms of averaged quantities. For example, the density of states near the Fermi level is lower on electron-doped side, and also the bare spin susceptibility at $(\pi,\pi)$ is stronger on the hole-doped side. Hence, we believe, that it is essential to resolve the wavevector-dependence of the interaction as done by the functional RG in order to see this trend.   
  
Next, we describe the dependence of the pairing instability on the initial interactions. Some data is shown in Fig \ref{phasediag}b). One clearly sees that the critical scale dramatically depends on $g_3$, with $g_3=0$, where we do not find any pairing instability at reasonable scales. This fact can be understood  already in the two-patch model and is also stated in Ref. \cite{Chubukov}. Our results show that the refined wavevector-dependence does not alter the situation, and that no other significant pairing instabilities occur if the extended $s$-wave channel is taken out by $g_3=0$. Hence a certain amount of pair scattering between electron- and hole-pockets is essential for a sizable $T_c$. Furthermore, we observe a repulsive interaction $g_1$ works against the pairing instability, but is otherwise not of central importance.  
  
Regarding the transition between SDW and extended $s$-wave pairing, we observe in the plots in Fig. \ref{flowres} for the coupling function that the SDW channel has only little overlap with the pairing channel. Hence, it is reasonable to assume that its flow and also the effective potential for the SDW order parameter will be only weakly affected by the pairing correlations at low $T$. Then the SDW order may reasonably well be described by a renormalized mean-field theory. 
 Meanfield studies of density-wave instabilities in imperfectly nested situations quite generally give a regime with first-order transitions, where above a critical doping $x$ the first-order transition line $T_t (x)$ extends from a tricritical point above the second order transition line $T_c(x)$ and drops to zero at some higher critical doping $x_c$\cite{gersch}. In our case, we have to keep in mind that the functional RG in this version is not sensitive with respect to first order- transitions\cite{gersch}. But as the channel coupling is rather weak, regarding the effective potential for the SDW order parameter, we expect a similar situation as in the meanfield picture. 
Then, the SDW phase would actually extend to higher doping than is depicted in Fig. \ref{phasediag}(a), with the shoulders given by first order SDW-to-metal transitions at temperatures above the pairing scale. As the energy difference between metal and superconductor is small, it is very likely that also the superconductivity-to-SDW transition at low $T$ is of first order as well. It will be interesting to check this scenario with more sophisticated calculations. Very recently, the  meanfield phase diagram for competing extended $s$-pairing and spin-density wave has been worked out in a two-pocket model for the pnictides\cite{vorontsov}, confirming the basic picture of a first order transition, with the possibility of an intermediate phase with incommensurate SDW phase coexisting with pairing for large differences of the energy scales for pairing and SDW. In our RG study we refrain from distinguishing between commensurate and slightly incommensurate order as the reduced wavevector resolution would not permit definite statements.
\section{Summary and discussion}
Our functional RG study of the four-band model clearly identifies SDW order near zero doping and extended $s$-wave pairing with a sign change of the pair amplitude between electron and hole pockets at nonzero electron and hole doping as the two leading instabilities. Hence, the basic picture for the instabilities by Chubukov et al.\cite{Chubukov} remains valid, even if one allows for more realistic non-spherical Fermi surfaces and more possibilities for variations of the coupling function within the Brillouin zone. The additional variation of the dominant interactions around the Fermi surface turn out to be mild, or irrelevant in the RG sense. According to our generalization, this two-patch physics connects smoothly to the more ambitious studies by the Berkeley group\cite{d_h_lee} who take into account the ab-initio band structure and start with multi-orbital interactions in real space. Compared to our study, this changes details of the band structure and adds additional wavevector dependence to the initial interactions. Yet, the leading instabilities turn out to be of the same type.  

The main reason for the extended $s$-wave pairing is identified as the $g_3$-like scattering between electron and hole pockets. This interaction also drives the SDW tendencies, as was emphasized recently by the Berkeley group\cite{zhai}. This important role of the particle-hole channel with momentum transfer $(\pi,\pi)$ is very similar to the weak coupling picture of the $d$-wave superconductivity in the one-band Hubbard model near half filling, which is possibly also relevant for the high-$T_c$ cuprates. In this sense, and if our modelling indeed captures the relevant physics, the pairing mechanism in cuprates and iron arsenides can be considered to be of the same type. The difference in the pairing symmetry is due to the different location of the Fermi surface parts responsible for the pairing instability. Furthermore, in the one-band Hubbard model on the square lattice, the competition and coupling between the various channels can be more intense\cite{hsfr}. 

Our study allows to obtain estimates for the instability scale as a function of doping and the critical doping where the SDW phase gives way to the pairing instability. Here, we find that the electron-doped side with smaller density of states near the Fermi surface shows the higher critical scale for pairing if we compare the same absolute value of the doping. This trend is also found experimentally. It shows that in anisotropic systems with electronically mediated pairing not only the averaged density of states decides on the energy scale for pairing. This scale can be a rather complicated function of the dispersion and interaction parameters. Notably, the observed trend cannot be explained by a higher degree of nesting on the electron-doped side. At least, if we take the bare particle-hole susceptibility at wavevector $(\pi,\pi)$ as a measure for the nesting, it turns out to be smaller on this side with the higher pairing scale.

Concerning the transition between SDW and extended $s$-wave pairing, the functional RG exhibits a continuous change of the instability with doping, with a smooth variation of the critical scale. As we do not include self-energy corrections, this scale should be considered as an upper bound for the breakdown of the metallic state, and a second order transition between SDW and extended $s$-wave pairing cannot be excluded rigorously. However, as the couping between the channels is rather weak, and as commensurate density-wave phases in imperfectly nested situation often end in first-order transitions, we have argued that a first-order transition between the SDW and the superconducting state is quite plausible.

Finally, we can compare our results with relatively simple bare interactions defined in the band picture with other approaches where the interactions are defined in the orbital real-space picture. We observe two differences, which indicate the influence of the 'matrix elements' set to unity in our study. One aspect is the question of competing pairing channels. The fRG with matrix elements\cite{d_h_lee} and also RPA-type calculations\cite{maier} find a $d$-wave component that is only slightly weaker than the extended $s$-wave channel.  In our calculations, the $d$-wave component is much weaker, and does not appear as a competing channel. Further, photoemission\cite{ding} and other theoretical studies\cite{maier} obtain a different magnitude of the pairing gap on the two hole pockets. Our results do not indicate strong differences between the pairing strengths of the 4 Fermi surfaces. Again, it is possible that this effect is due to the matrix elements that give additional structure to the bare interactions.

\ack
We thank Jutta Ortloff, Thomas Maier, D.J. Scalapino and I. Eremin for helpful discussions and the DFG research unit FOR 538 for support.



\end{document}